\documentclass[a4paper, 9pt, twocolumn]{IEEEtran}      

\usepackage{amsmath,amssymb,enumerate,soul}
\usepackage{epstopdf}
\usepackage{times} 
\usepackage[dvips]{graphicx}
\usepackage{psfrag}
\usepackage{multirow} 
\usepackage{cite} 
\usepackage{float}
\usepackage[x11names]{xcolor}
\usepackage{colortbl}

\usepackage{amsthm}
\theoremstyle{plain}
\newtheorem{ass}{Assumption}
\newtheorem{thm}{Theorem}
\newtheorem{props}{Proposition}
\newtheorem{cor}{Corollary}

\theoremstyle{definition}
\newtheorem{exmple}{Example}

\newtheoremstyle{cited}%
  {3pt}
  {3pt}
  {\itshape}
  {}
  {\bfseries}
  {.}
  {.5em}
  {\thmname{#1} \thmnumber{#2} \normalfont(\thmnote{\normalfont#3})}
\theoremstyle{cited}
\newtheorem{lemma}{Lemma}

\newcommand{\arctanh}{\operatorname{arctanh}}
\newcommand{\dom}{\ensuremath{\text{dom}\,}}
\newcommand{\avg}{\ensuremath{\text{avg}\,}}

\newcommand{\R}[2]{\ensuremath{\mathbb{R}^{#1}_{#2}}}
\newcommand{\Zp}{\ensuremath{\mathbb{Z}_{\geq 0}}}
\newcommand{\K}{\ensuremath{\mathcal{K}}}
\newcommand{\Kinf}{\ensuremath{\mathcal{K}_{\infty}}}
\newcommand{\KL}{\ensuremath{\mathcal{KL}}}

\graphicspath{{Figures/}}

\title{\LARGE \bf
Stabilization of nonlinear systems using event-triggered output feedback controllers
}

\author{Mahmoud Abdelrahim, Romain Postoyan, Jamal Daafouz and Dragan Ne\v{s}i\'c
\thanks{M. Abdelrahim, R. Postoyan and J. Daafouz are with the Universit\'{e} de Lorraine, CRAN, UMR 7039 and the CNRS, CRAN, UMR 7039, France {\tt\small \{othmanab1,romain.postoyan,jamal.daafouz\}@} {\tt\small univ-lorraine.fr}. Their work is partially supported by the ANR under the grant COMPACS (ANR-13-BS03-0004-02).}
\thanks{D. Ne\v{s}i\'c is with the Department of Electrical and Electronic Engineering, the University of Melbourne, Parkville, VIC 3010, Australia {\tt\small dnesic@unimelb.edu.au}. His work is supported by the Australian Research Council under the Discovery Projects and Future Fellowship schemes.}
}

\begin{document}

\maketitle

\begin{abstract}
The objective is to design output feedback event-triggered controllers to stabilize a class of nonlinear systems. One of the main difficulties of the problem is to ensure the existence of a minimum amount of time between two consecutive transmissions, which is essential in practice. We solve this issue by combining techniques from event-triggered and time-triggered control. The idea is to turn on the event-triggering mechanism only after a fixed amount of time has elapsed since the last transmission. This time is computed based on results on the stabilization of time-driven sampled-data systems. The overall strategy ensures an asymptotic stability property for the closed-loop system. The results are proved to be applicable to linear time-invariant (LTI) systems as a particular case.
\end{abstract}

\section{Introduction}
Networked control systems (NCS) are systems in which the communication between the plant and the controller occurs through a shared digital channel. Since the network has a limited bandwidth and is typically used by other tasks, it is essential to develop communication-aware control strategies. Event-triggered control is a relevant paradigm in this context as it adapts transmissions to the current state of the plant, see \textit{e.g.} \cite{Arzen1999simple, Astrom2002comparison, Tabuada2007event, Romain2011unifying, Heemels2012Introduction} and the references therein. In that way, transmissions only occur when it is needed according to the control objectives.

A fundamental issue in the implementation of event-triggered controllers is to ensure the existence of a minimum amount of time between two consecutive transmissions to respect hardware limitations. This task becomes particularly challenging when we have to design the controller using only an output of the system and not the full state vector (see \cite{Donkers2012output}), in particular when we aim to guarantee asymptotic stability properties. To the best of our knowledge, this problem has been first addressed in \cite{Kofman2006level} and then in \cite{Donkers2012output,Lehmann2011event, Peng2013output,Tallapragada2012event-CDC,Forni2014event,Zhang2013event,Meng2014event} for LTI systems and in \cite{Yu2012event} for nonlinear systems.

In this paper, we design output feedback event-triggered controllers for nonlinear systems which guarantee a (global) asymptotic stability property and the existence of a uniform strictly positive lower bound on the inter-transmission times. The proposed strategy combines the event-triggering condition of \cite{Tabuada2007event} adapted to output measurements and the results on time-driven sampled-data systems in \cite{Nesic2009explicit}. Indeed, the event-triggering condition is only (continuously) evaluated after $T$ units of times have elapsed since the last transmission, where $T$ corresponds to the \textit{maximum allowable sampling period} (MASP) given by \cite{Nesic2009explicit}. This two-step procedure is justified by the fact that the adaption of the event-triggering condition of \cite{Tabuada2007event} to output feedback on its own can lead to Zeno phenomenon (see \cite{Donkers2012output}). Although the rationale is intuitive, the analysis is not trivial as we show in the paper. This type of triggering rule has been used in \cite{Abdelrahim2013Event} to stabilize nonlinear singularly perturbed systems under a different set of assumptions. Similar approaches have been followed in \cite{Forni2014event}, \cite{Mazo2011decentralized}, \cite{Wang2012asynchronous} to enforce a lower bound on the inter-transmission times in different contexts, mainly for linear systems. Note that the idea of enforcing a given time between two jumps is linked to time regularization techniques, see \cite{Johansson1999simulation}.

Our results rely on similar assumptions as in \cite{Nesic2009explicit} which allow us to derive both local and global results. These conditions are shown to be always verified by LTI systems that are stabilizable and detectable, in which case these are reformulated as a linear matrix inequality (LMI). Contrary to \cite{Forni2014event}, the approach is applicable to nonlinear systems and the output feedback law is not necessarily based on an observer. Compared to \cite{Yu2012event}, we rely on a different set of assumptions and we conclude a different stability property. In addition, we show that our results are applicable to any LTI systems that are stabilizable and detectable, which is a priori not the case of \cite{Yu2012event}. Furthermore, we apply our results to Lorenz model, which is nonlinear and which does not satisfy the conditions of \cite{Yu2012event}. Unlike \cite{Tallapragada2012event-CDC}, where LTI systems have been studied, we do not necessarily consider observer-based output feedbacks and the triggering condition does not a priori rely on estimates of the unmeasured states. The latter has the advantage to lighten the implementation since the triggering mechanism only needs to have access to the output of the plant, and not the controller variable. Finally, in the particular case of LTI systems, we conclude a global asymptotic stability property as opposed to ultimate boundedness in \cite{Donkers2012output}. It has to be noted that the event-triggering mechanism that we propose is different from the periodic event-triggered control (PETC) paradigm, see \textit{e.g.} \cite{Heemels2013PETC}, \cite{Romain2013periodic}, where the triggering condition is verified only at some periodic sampling instants. In our case, the triggering mechanism is \textit{continuously} evaluated, once $T$ units of time have elapsed since the last transmission. The first results of this work have been presented in \cite{Abdelrahim2014stabilization}. In comparison to our previous work: we provide all the proofs of the results; we show how the proposed technique can be fruitfully employed in the context of state feedback control as a special case, to directly tune the lower bound on the inter-transmission times; we apply the results on a different physical nonlinear example to better motivate our results and we compare our proposed triggering mechanism with the existing results on a linear output feedback example.


\section{Preliminaries} \label{sec: preliminaries}
Let $\R{}{} := (-\infty,\infty)$, $\R{}{\geq 0} := [0,\infty)$ and $\Zp := \{ 0, 1, 2, . . \}$. A continuous function $\gamma: \R{}{\geq 0} \rightarrow \R{}{\geq 0}$ is of class $\K$ if it is zero at zero, strictly increasing, and it is of class $\Kinf$ if in addition $\gamma(s) \rightarrow \infty$ as $s \rightarrow \infty$. A continuous function $\gamma: \R{}{\geq 0} \times \R{}{\geq 0} \rightarrow \R{}{\geq 0}$ is of class $\KL$ if for each $t \in \R{}{\geq 0}$, $\gamma(.,t)$ is of class $\K$, and, for each $s \in \R{}{\geq 0}$, $\gamma(s,.)$ is decreasing to zero. We denote the minimum and maximum eigenvalues of the symmetric matrix $A$ as $\lambda_{\min}(A)$ and $\lambda_{\max}(A)$, respectively. We write $A^{T}$ to denote the transpose of $A$. We use $\mathbb{I}_{n}$ to denote the identity matrix of dimension $n$. We write $(x,y)$ to represent the vector $[x^{T}, y^{T}]^{T}$ for $x \in \R{n}{}$ and $y \in \R{m}{}$. For a vector $x\in\R{n}{}$, we denote by $|x|:=\sqrt{x^{T}x}$ its Euclidean norm and for a matrix $A\in\R{n\times m}{}$, we denote by $|A| := \sqrt{\lambda_{\max}(A^{T}A)}$. We will consider locally Lipschitz Lyapunov functions (that are not necessarily differentiable everywhere), therefore we will use the generalized directional derivative of Clarke which is defined as follows. For a locally Lipschitz function $V: \R{n}{} \rightarrow \R{}{\geq 0}$ and a vector $\upsilon \in \R{n}{}$, $V^{\circ}(x;\upsilon) := \lim \sup_{h \to 0^{+}, \, y \to x}(V(y+h\upsilon) - V(y))/h$. For a continuously differentiable function $V$, $V^{\circ}(x;\upsilon)$ reduces to the standard directional derivative $\langle\nabla V(x), \upsilon\rangle$, where $\nabla V(x)$ is the (classical) gradient. We will invoke the following result, see Lemma II.1 in \cite{Liberzon2012Lyapunov}.

\begin{lemma}[Lemma II.1 \cite{Liberzon2012Lyapunov}] \label{lma: clarke}
Consider two functions $U_{1}: \R{n}{} \rightarrow \R{}{}$ and $U_{2}: \R{n}{} \rightarrow \R{}{}$ that have well-defined Clarke derivatives for all $x\in\R{n}{}$ and $\upsilon \in \R{n}{}$. Introduce three sets $A:=\{ x: U_{1}(x) > U_{2}(x) \}$, $B:=\{ x: U_{1}(x) < U_{2}(x) \}$, $\Gamma:= \{ x: U_{1}(x) = U_{2}(x) \}$. Then, for any $\upsilon \in \R{n}{}$, the function $U(x) := \max\{U_{1}(x), U_{2}(x)\}$ satisfies $U^{\circ}(x; \upsilon) = U_{1}^{\circ}(x;\upsilon)$ for all $x\in A$, $U^{\circ}(x; \upsilon) = U_{2}^{\circ}(x;\upsilon)$ for all $x\in B$ and $U^{\circ}(x; \upsilon) \leq \max\{U_{1}^{\circ}(x;\upsilon), U_{2}^{\circ}(x;\upsilon)\}$ for all $x\in \Gamma$. \hfill $\Box$
\end{lemma}

In this paper, we consider hybrid systems of the following form using the formalism of \cite{Teel}
\begin{equation}\label{eq: Prelim-hybrid-model}
  \dot{x} = F(x) \hspace{10pt} x\in C, \hspace{20pt} x^{+} = G(x) \hspace{10pt} x\in D,
\end{equation}
where $x\in \R{n}{}$ is the state, $F$ is the flow map, $C$ is the flow set, $G$ is the jump map and $D$ is the jump set. The vector fields $F$ and $G$ are assumed to be continuous and the sets $C$ and $D$ are closed. The solutions to system (\ref{eq: Prelim-hybrid-model}) are defined on so-called hybrid time domains. A set $E \subset \R{}{\geq0}\times \Zp$ is called a \textit{compact hybrid time domain} if $E =  \displaystyle \underset{j\in\{0,...,J-1\}}{\cup}([t_{j}, t_{j+1}], j)$ for some finite sequence of times $0=t_{0}\leq t_{1} \leq ... \leq t_{J}$ and it is a \textit{hybrid time domain} if for all $(T,J)\in E, E \cap ([0,T]\times \{0,1,...,J\})$ is a compact hybrid time domain. A function $\phi:E\rightarrow\R{n}{}$ is a hybrid arc if $E$ is a hybrid time domain and if for each $j\in\Zp, t \mapsto \phi(t,j)$ is locally absolutely continuous on $I^{j} := \{ t: (t,j)\in E \}$. A hybrid arc $\phi$ is a solution to system (\ref{eq: Prelim-hybrid-model}) if: (i) $\phi(0,0)\in C\cup D$; (ii) for any $j\in \Zp$, $\phi(t,j) \in C$ and $\dot{\phi}(t,j) = F(\phi(t,j))$ for almost all $t \in I^{j}$; (iii) for every $(t,j)\in \dom \phi$ such that $(t,j+1)\in \dom \phi$, $\phi(t,j) \in D$ and $\phi(t,j+1) = G(\phi(t,j))$. A solution $\phi$ to system (\ref{eq: Prelim-hybrid-model}) is \textit{maximal} if it cannot be extended, \textit{complete} if its domain, $\dom \phi$, is unbounded, and it is \textit{Zeno} if it is complete and $\sup_{t}\dom \phi< \infty$.

\section{Problem statement} \label{sec: problem-statement}
\begin{figure}[h!]
\centering \scriptsize
\psfrag{Plant}[][][1]{Plant}
\psfrag{ETC}[][][1]{Event-triggering}
\psfrag{mechanism}[][][1]{mechanism}
\psfrag{Controller}[][][1]{Controller}
\psfrag{yt}[][][1]{$y(t)$}
\psfrag{yk}[][][1]{$y(t_{i})$}
\psfrag{ut}[][][1]{$u(t)$}
\psfrag{uk}[][][1]{$u(t_{i})$}
\includegraphics[width=7cm,height=3cm]{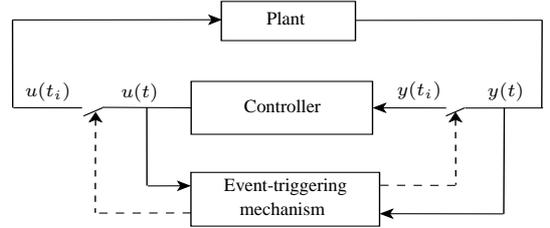}
\caption{Event-triggered control schematic \cite{Donkers2012output}}\label{fig:output-controller}
\end{figure}

Consider the nonlinear plant model
\begin{equation} \label{eq: xdot_p-y}
\begin{array}{lllll}
  \dot{x}_{p} &=& f_{p}(x_{p},u), \hspace{25pt} y &=& g_{p}(x_{p}),
\end{array}
\end{equation}
where $x_{p} \in \R{n_{p}}{}$ is the plant state, $u \in \R{n_{u}}{}$ is the control input, $y\in\R{n_{y}}{}$ is the measured output of the plant and we focus on general dynamic controllers of the form
\begin{equation}\label{eq: xdot_c-u}
\begin{array}{lllll}
  \dot{x}_{c} &=& f_{c}(x_{c}, y), \hspace{25pt}  u &=& g_{c}(x_{c}, y),
\end{array}
\end{equation}
where $x_{c} \in \R{n_{c}}{}$ is the controller state. We emphasize that the $x_{c}$-system is not necessarily an observer. Moreover, (\ref{eq: xdot_c-u}) captures static feedbacks as a particular case by setting $u=g_{c}(y)$. We follow an emulation approach in this paper. Hence, we assume that the controller (\ref{eq: xdot_c-u}) renders the origin of system (\ref{eq: xdot_p-y}) globally asymptotically stable in the absence of network. Afterwards, we take into account the communication constraints and we synthesize the triggering condition. In particular, we consider the scenario where controller (\ref{eq: xdot_c-u}) communicates with the plant via a digital channel. Hence, the plant output and the control input are sent only at transmission instants $t_{i}, i\in \Zp$. We are interested in an event-triggered implementation in the sense that the sequence of transmission instants is determined by a criterion based on the output measurement, see Figure \ref{fig:output-controller}.
At each transmission instant, the plant output is sent to the controller which computes a new control input that is instantaneously transmitted to the plant. We assume that this process is performed in a synchronous manner and we ignore the computation times and the possible transmission delays. In that way, we obtain
\begin{equation}
\left.
  \begin{array}{rcll}
    \dot{x}_{p} &=& f_{p}(x_{p},\hat{u}) &\hspace{30pt}  t\in [t_{i}, t_{i+1}]\\
    \dot{x}_{c} &=& f_{c}(x_{c}, \hat{y}) &\hspace{30pt}  t\in [t_{i}, t_{i+1}]\\
    u &=& g_{c}(x_{c}, \hat{y}) \\
    \dot{\hat{y}} &=& 0 &\hspace{30pt}  t\in [t_{i}, t_{i+1}]\\
    \dot{\hat{u}} &=& 0 &\hspace{30pt}  t\in [t_{i}, t_{i+1}] \\
    \hat{y}(t_{i}^{+}) &=& y(t_{i}) \\
    \hat{u}(t_{i}^{+}) &=& u(t_{i}),
  \end{array}
\right \}
\end{equation}
\noindent where $\hat{y}$ and $\hat{u}$ respectively denote the last transmitted values of the plant output and the control input. We assume that zero-order-hold devices are used to generate the sampled values $\hat{y}$ and $\hat{u}$, which leads to  $\dot{\hat{y}} = 0$ and $\dot{\hat{u}} = 0$. We introduce the network-induced error $e := (e_{y}, e_{u}) \in \R{n_{e}}{}$, where $e_{y} := \hat{y} - y$ and $e_{u} := \hat{u} - u$ which are reset to $0$ at each transmission instant.

We model the event-triggered control system using the hybrid formalism of \cite{Teel} as in \cite{Donkers2012output}, \cite{Forni2014event}, \cite{Romain2011unifying}, for which a jump corresponds to a transmission. In that way, the system is modeled as
\begin{equation} \label{eq: hybrid-model}
\begin{array}{rllll}
\left(\begin{array}{c} \dot{x}\\ \dot{e}\\ \dot{\tau}\end{array} \right) &=& \left(\begin{array}{c} f(x,e)\\ g(x,e)\\ 1\end{array} \right) &\hspace{15pt}(x,e,\tau)\in C, \\[15pt]
  \left(\begin{array}{c} x^{+}\\ e^{+}\\ \tau^{+}\end{array} \right) &=& \left(\begin{array}{c} x\\ 0\\ 0\end{array} \right) &\hspace{15pt} (x,e,\tau) \in D,
\end{array}
\end{equation}
where $x:=(x_{p}, x_{c}) \in \R{n_{x}}{}$ and $\tau \in \R{}{\geq 0}$ is a clock variable which describes the time elapsed since the last jump,
$f(x,e) = ( f_{p}(x_{p},g_{c}(x_{c}, y + e_{y}) + e_{u}),\, f_{c}(x_{c}, y + e_{y}))$  and \\
$g(x,e) = (-\frac{\partial}{\partial x_{p}}g_{p}(x_{p})f_{p}(x_{p},g_{c}(x_{c}, y + e_{y}) + e_{u}),\, - \frac{\partial}{\partial x_{c}}g_{c}(x_{c}, y + e_{y})f_{c}(x_{c}, y + e_{y}) )$.
The flow and jump sets of (\ref{eq: hybrid-model}) are defined according to the triggering condition we will define. As long as the triggering condition is not violated, the system flows on $C$ and a jump occurs when the state enters in $D$. When $(x,e,\tau) \in C \cap D$, the solution may flow only if flowing keeps $(x,e,\tau)$ in $C$, otherwise the system experiences a jump. The functions $f$ and $g$ are assumed to be continuous and the sets $C$ and $D$ will be closed (which ensure that system (\ref{eq: hybrid-model}) is well-posed, see Chapter 6 in \cite{Teel}).

The main objective of this paper is to design the flow and the jump sets of system (\ref{eq: hybrid-model}), \textit{i.e.} the triggering condition, to ensure a (global) asymptotic stability property for system (\ref{eq: hybrid-model}).

\section{Main results}\label{sec: main-results}
We first present the conditions that we impose on system (\ref{eq: hybrid-model}), then we present the triggering technique and finally we state the main result. We make the following assumption on system (\ref{eq: hybrid-model}), which is inspired by \cite{Nesic2009explicit}.

\begin{ass} \label{ass-clock-out}
There exist $\Delta_{x}, \Delta_{e}>0$, locally Lipschitz positive definite functions $V: \R{n_{x}}{}\rightarrow\R{}{\geq 0}$ and $W: \R{n_{e}}{}\rightarrow\R{}{\geq 0}$, continuous function $H: \R{n_{x}}{}\rightarrow\R{}{\geq 0}$, real numbers $\gamma, L\geq 0$, $\underline{\alpha}, \overline{\alpha} \in \Kinf$ and continuous, positive definite functions $\delta: \R{n_{y}}{}\rightarrow\R{}{\geq 0}$ and $\alpha: \R{}{\geq 0}\rightarrow\R{}{\geq 0}$ such that, for all $x\in\R{n_{x}}{}$
\begin{align} \label{Vx-bounds-out}
  \underline{\alpha}(|x|) \leq V(x) \leq \overline{\alpha}(|x|),
\end{align}
for all $|e| \leq \Delta_{e}$ and almost all $|x| \leq \Delta_{x}$
\begin{align} \label{V-dot-out}
  \langle\nabla V(x), f(x,e)\rangle \leq -\alpha(|x|) - H^{2}(x) - \delta(y) + \gamma^{2} W^{2}(e)
\end{align}
and for all $|x| \leq \Delta_{x}$ and almost all $|e| \leq \Delta_{e}$
\begin{align}\label{W-dot-out}
  \langle\nabla W(e), g(x,e)\rangle \leq LW(e) + H(x).
\end{align}
We say that Assumption \ref{ass-clock-out} holds globally if (\ref{V-dot-out}) and (\ref{W-dot-out}) hold for almost all $x\in\R{n_{x}}{}$ and $e\in\R{n_{e}}{}$.
\hspace*{\fill} $\Box$
\end{ass}

Conditions (\ref{Vx-bounds-out})-(\ref{V-dot-out}) imply that the system $\dot{x} = f(x,e)$ is $\mathcal{L}_{2}$-gain stable from $W$ to $(H,\sqrt{\delta})$. This property can be analysed by investigating the robustness property of the closed-loop system (\ref{eq: xdot_p-y})-(\ref{eq: xdot_c-u}) with respect to input and/or output measurement errors in the absence of sampling. Note that, since $W$ is positive definite and continuous (since it is locally Lipschitz), there exists $\chi\in\Kinf$ such that $W(e) \leq \chi(|e|)$ (according to Lemma 4.3 in \cite{Khalil}) and hence (\ref{Vx-bounds-out}), (\ref{V-dot-out}) imply that the system $\dot{x}=f(x,e)$ is input-to-state stable (ISS). We also assume an exponential growth condition of the $e$-system on flows in (\ref{W-dot-out}) which is similarly used in \cite{Nesic2009explicit}. We provide an example of nonlinear system which verifies Assumption \ref{ass-clock-out} at the end of this section.

Under Assumption \ref{ass-clock-out}, the adaptation of the idea of \cite{Tabuada2007event} leads to a triggering condition of the form $\gamma^{2} W^{2}(e) \leq \delta(y)$. The problem is that Zeno phenomenon may occur with this type of triggering conditions. Indeed, when $y=0$, an infinite number of jumps occurs for any value of $x$ such that $g_{p}(x_{p})=0$. In \cite{Donkers2012output}, this issue is overcome by adding a constant to the triggering condition, which would lead to $\gamma^{2} W^{2}(e) \leq \delta(y) + \varepsilon$ here for $\varepsilon > 0$, from which we can derive a practical stability property. The event-triggered mechanism that we propose allows us to guarantee an asymptotic stability property for the closed-loop while ensuring that the inter-transmission times are lower bounded by a strictly positive constant. The idea is to evaluate the event-triggering condition only after $T$ units have elapsed since the last transmission, where $T$ corresponds to the MASP given by \cite{Nesic2009explicit}. In that way, we allow the user to directly tune the minimum inter-jump interval, up to a certain extent as explained in the following. We thus redesign the triggering condition as follows
\begin{equation} \label{eq:tig-condn}
  \gamma^{2} W^{2}(e) \leq \delta(y) \text{ or } \tau \in [0,T],
\end{equation}
where we recall that $\tau\in \R{}{\geq 0}$ is the clock variable introduced in (\ref{eq: hybrid-model}). Consequently, the flow and jump sets of system (\ref{eq: hybrid-model}) are
\begin{equation}\label{flow-jump-sets-out}
\arraycolsep=1pt\def\arraystretch{1.3}
\begin{array}{lcll}
C &=& \Big\{(x,e,\tau): &\gamma^{2} W^{2}(e) \leq \delta(y) \text{ or } \tau \in [0,T]\Big\} \\[6pt]
D &=& \Big\{(x,e,\tau): &\Big(\gamma^{2} W^{2}(e) = \delta(y) \text{ and } \tau \geq T\Big) \text{ or } \\[6pt]
& & & \Big(\gamma^{2} W^{2}(e) \geq \delta(y) \text{ and } \tau = T\Big) \Big\}.
\end{array}
\end{equation}
Hence, the inter-jump times are uniformly lower bounded by $T$. This constant is selected such that $T < \mathcal{T}(\gamma, L)$, where
\begin{equation} \label{T-phi}
\mathcal{T}(\gamma, L) := \left\{
    \begin{array}{ll}
    \frac{1}{Lr}\arctan(r)& \hspace{10pt} \gamma > L \\
    \frac{1}{L}& \hspace{10pt} \gamma = L \\
    \frac{1}{Lr}\arctanh(r)& \hspace{10pt} \gamma < L
    \end{array}
    \right.
\end{equation}
with $r := \sqrt{\left|(\frac{\gamma}{L})^{2} - 1\right|}$ and $L,\gamma$ come from Assumption \ref{ass-clock-out} as in \cite{Nesic2009explicit}. We are ready to state the main result.

\begin{thm} \label{thm: clock-output}
Suppose that Assumption \ref{ass-clock-out} holds and consider system (\ref{eq: hybrid-model}) with the flow and jump sets (\ref{flow-jump-sets-out}), where the constant $T$ is such that $T \in (0,\mathcal{T}(\gamma, L))$. There exist $\Delta> 0$ and $\beta \in \KL$ such that any solution $\phi = (\phi_{x},\phi_{e},\phi_{\tau})$ with $|(\phi_{x}(0,0), \phi_{e}(0,0))|\leq\Delta$  satisfies
\begin{equation} \label{eq-thm-clock-out}
|\phi_{x}(t,j)| \leq \beta(|(\phi_{x}(0,0), \phi_{e}(0,0))|, t + j) \hspace{10pt} \forall(t,j) \in \dom \phi,
\end{equation}
furthermore, if $\phi$ is maximal, then it is complete. If Assumption \ref{ass-clock-out} holds globally, then (\ref{eq-thm-clock-out}) holds globally.
\hfill $\Box$
\end{thm}

\begin{exmple}\label{sec: illutrative-example}
Consider the controlled Lorenz equations which model fluid convection \cite{Wan95nonlinear}, $\dot{x}_{1} = -a x_{1} + a x_{2}$, $\dot{x}_{2} = bx_{1} - x_{2} - x_{1}x_{3} + u$, $\dot{x}_{3} = x_{1}x_{2} - cx_{3}$ and $y = x_{1}$, where $a,b,c > 0$. The static output feedback law $u = -(\frac{p_{1}}{p_{2}}a + b)x_{1}$, where $p_{1}, p_{2}>0$, globally stabilizes the origin. This can be proved by using the quadratic Lyapunov function $V(x) = p_{1}x_{1}^{2} + p_{2}x_{2}^{2} + p_{2}x_{3}^{2}$, which verify condition (\ref{Vx-bounds-out}) with $\underline{\alpha}(|x|)=\min\{p_{1},p_{2}\}|x|^{2}$ and $\overline{\alpha}(|x|)=\max\{p_{1},p_{2}\}|x|^{2}$. We take into account the network-induced error $e = \hat{y} - y$ (it is not necessary to consider the error in $u$ as the controller is static) and we select $W(e) = |e|$. Hence, condition (\ref{W-dot-out}) is satisfied with $L = 0$, $H(x) = a(|x_{1}| + |x_{2}|)$. By taking $p_{1}>1$ and $p_{2}>2a$, condition (\ref{V-dot-out}) holds with $\alpha(|x|) = \min\{a(p_{1} - 1), (p_{2} - 2a), 2p_{2}c\}|x|^{2}$, $\delta(y) = a(p_{1} - 1)y^{2}$ and $\gamma^{2} = p_{2}(\frac{p_{1}}{p_{2}}a + c)^{2}$. For the parameter values $a=10, b= 28, c=8/3$ used in \cite{Wan95nonlinear}, we set $p_{1} = 2$, $p_{2} = 3a$ and we obtain $T = 0.01$. We note that the results in \cite{Yu2012event} are not applicable to this system because condition (3) of Proposition 1 in \cite{Yu2012event} does not hold. \hfill $\Box$
\end{exmple}


\section{Linear systems}\label{sec: application-linear}
We now focus on the particular case of linear systems. Consider the LTI plant model
\begin{equation}\label{plant-linear-ex-out}
\dot{x}_{p} = A_{p}x_{p} + B_{p}u, \hspace{30pt} y = C_{p}x_{p},
\end{equation}
where $x_{p}\in \R{n_{p}}{}$, $u\in \R{n_{u}}{}$, $y\in \R{n_{y}}{}$ and $A_{p}, B_{p}, C_{p}$ are matrices of appropriate dimensions. We design the following dynamic controller to stabilize (\ref{plant-linear-ex-out}) in the absence of sampling
\begin{equation}\label{controller-linear-ex-out}
\dot{x}_{c} = A_{c}x_{c} + B_{c}y, \hspace{30pt} u = C_{c}x_{c} + D_{c}y,
\end{equation}
where $x_{c} \in \R{n_{c}}{}$ and $A_{c}, B_{c}, C_{c}, D_{c}$ are matrices of appropriate dimensions. Afterwards, we take into account the communication constraints. Then, the hybrid model (\ref{eq: hybrid-model}) is
\begin{equation} \label{eq: hybrid-model-linear}
\begin{array}{rllllll}
\left(\begin{array}{c} \dot{x}\\ \dot{e}\\ \dot{\tau}\end{array} \right) &=& \left(\begin{array}{c} \mathcal{A}_{1}x + \mathcal{B}_{1}e\\ \mathcal{A}_{2}x +  \mathcal{B}_{2}e\\ 1\end{array} \right) &\hspace{10pt} (x,e,\tau)\in C \\[15pt]
\left(\begin{array}{c} x^{+}\\ e^{+}\\ \tau^{+}\end{array} \right) &=& \left(\begin{array}{c} x\\ 0\\ 0\end{array} \right) &\hspace{10pt} (x,e,\tau) \in D,
\end{array}
\end{equation}
\begin{flushleft}
\linespread{2.5}\selectfont
where $\mathcal{A}_{1} := \left(\begin{smallmatrix} A_{p} + B_{p}D_{c}C_{p} &\hspace{6pt} B_{p}C_{c}\\[6pt] B_{c}C_{p} &\hspace{6pt} A_{c} \end{smallmatrix} \right)$, $\mathcal{B}_{1} :=  \left(\begin{smallmatrix} B_{p}D_{c} &\hspace{6pt} B_{p}\\[6pt] B_{c} &\hspace{6pt} 0 \end{smallmatrix} \right)$, $\mathcal{A}_{2} := \left(\begin{smallmatrix} - C_{p}(A_{p} + B_{p}D_{c}C_{p}) &\hspace{6pt} - C_{p}B_{p}C_{c}\\[6pt] - C_{c}B_{c}C_{p} &\hspace{6pt} - C_{c}A_{c} \end{smallmatrix} \right)$ and $\mathcal{B}_{2} := \left(\begin{smallmatrix} -C_{p}B_{p}D_{c} &\hspace{6pt} - C_{p}B_{p}\\[6pt] -C_{c}B_{c} &\hspace{6pt} 0 \end{smallmatrix} \right)$.
\end{flushleft}
\vspace{6pt}
We obtain the following result.
\begin{props}\label{prop: linear-systems}
Consider system (\ref{eq: hybrid-model-linear}). Suppose that there exist $\varepsilon_{1}, \varepsilon_{2}, \mu > 0$ and a positive definite symmetric real matrix $P$ such that
\begin{equation}\label{eq: prop-LMI}
\left(\begin{array}{cccc} \mathcal{A}_{1}^{T}P + P\mathcal{A}_{1}+ \mathcal{A}_{2}^{T}\mathcal{A}_{2} + \varepsilon_{1}\overline{C}_{p}^{T}\overline{C}_{p} + \varepsilon_{2}\mathbb{I}_{n_{x}}  &\hspace{12pt} P\mathcal{B}_{1} \\ \mathcal{B}_{1}^{T}P & -\mu\mathbb{I}_{n_{e}} \end{array} \right)\leq 0,
\end{equation}
where $\overline{C}_{p} = [C_{p} \hspace{12pt} 0]$. Then Assumption \ref{ass-clock-out} holds with
\begin{equation*}
\begin{array}{llllllllllll}
V(x) &=& x^{T}Px, & \underline{\alpha}(|x|) &=& \lambda_{\min}(P)|x|^{2}, \\
\overline{\alpha}(|x|) &=& \lambda_{\max}(P)|x|^{2}, & W(e) &=& |e|, \\
H(x) &=& |\mathcal{A}_{2}x|, & L &=& |\mathcal{B}_{2}|, \\
\gamma &=& \sqrt{\mu}, & \alpha(|x|) &=& \varepsilon_{2}|x|^{2}, \\
\delta(y) &=& \varepsilon_{1}|y|^{2}.
\end{array}
\end{equation*}
\hspace*{\fill} $\Box$
\end{props}
\noindent Proposition \ref{prop: linear-systems} provides a sufficient condition, namely (\ref{eq: prop-LMI}), for the verification of Assumption \ref{ass-clock-out}, which thus allows us to apply the results of Section \ref{sec: main-results}. It has to be noted that the LMI (\ref{eq: prop-LMI}) can always be satisfied when system (\ref{plant-linear-ex-out}) is stabilizable and detectable. Indeed, in this case, we can select the controller (\ref{controller-linear-ex-out}) such that $ \mathcal{A}_1$ is Hurwitz.  Noting that (\ref{eq: prop-LMI}) is equivalent to the following inequalities, by using the Schur complement of (\ref{eq: prop-LMI}) (see Section A.5.5 in \cite{Boyd}), $ \mathcal{A}_{1}^{T}P + P\mathcal{A}_{1}+ \mathcal{A}_{2}^{T}\mathcal{A}_{2} + \varepsilon_{1}\overline{C}_{p}^{T}\overline{C}_{p} + \varepsilon_{2}\mathbb{I}_{n_{x}} + \frac{1}{\mu}P\mathcal{B}_{1}\mathcal{B}_{1}^{T}P \leq 0$. We see that we can select the matrix $P$ such that $\mathcal{A}_{1}^{T}P + P\mathcal{A}_{1}+ \mathcal{A}_{2}^{T}\mathcal{A}_{2} + \varepsilon_{1}\overline{C}_{p}^{T}\overline{C}_{p} + \varepsilon_{2}\mathbb{I}_{n_{x}}$ is negative definite. It then suffices to choose $\mu$ sufficiently large to ensure the last inequality.

\begin{exmple}\label{sec: illutrative-example}
We apply the result in this section to Example 2 in \cite{Donkers2012output}. We obtain $L = 4$ and we obtain the values $\varepsilon_{1} = 1.5839$, $\varepsilon_{2}=13.9969$, $\gamma = 89.9666$ by solving the LMI (\ref{eq: prop-LMI}) using the SEDUMI solver with the YALMIP interface. The guaranteed minimum inter-transmission time is $T = 0.017$, by using (\ref{T-phi}). Table \ref{tbl:linear-out-emulation} provides the minimum and the average inter-transmission times, respectively denoted as $\tau_{\min}$ and $\tau_{\avg}$, for 100 randomly distributed initial conditions such that $|(x(0,0),e(0,0))|\leq 25$ and $\tau(0,0)=0$.  The provided values of $\tau_{\avg}$ in Table \ref{tbl:linear-out-emulation} indicates that the generated amount of transmissions by our proposed triggering mechanism is approximately 100 times less than the amount given by \cite{Donkers2012output}. Moreover, the stability property achieved in \cite{Donkers2012output} is a practical stability property, while we ensure a global asymptotic stability property. We note that the results in \cite{Yu2012event} are not applicable to this system because condition (3) of Proposition 1 in \cite{Yu2012event} is not satisfied. We believe that the comparison with \cite{Tallapragada2012event-CDC} is not relevant since the triggering mechanism is different and the dynamic controller in \cite{Tallapragada2012event-CDC} is based on an observer.
\hfill $\Box$

\begin{table}[H]
\begin{center}
\begin{tabular}{ c | c | c }
            & Donkers \& Heemels \cite{Donkers2012output}  & Our proposed mechanism \\ \hline
  Guaranteed & \multirow{2}{*}{6.5$\times 10^{-9}$}  &  \multirow{2}{*}{0.017}      \\
  lower bound & & \\ \hline
   $\tau_{\min}$ & $4.8055\times10^{-6}$  &  0.017 \\ \hline
   $\tau_{\avg}$ & $2.2905\times10^{-4}$  & 0.0202
  \end{tabular}
  \caption{Simulation results for 100 randomly distributed initial conditions such that $|(x(0,0),e(0,0))|\leq 25$ and $\tau(0,0)=0$ for a simulation time of 20 seconds.}
  \label{tbl:linear-out-emulation}
\end{center}
\end{table}
\end{exmple}

\section{State feedback controllers}\label{sec: linear-state-feedback}
The technique proposed in Section \ref{sec: main-results} is also relevant in the context of state feedback control, \textit{i.e.} when $y = x$, as the constant $T$ in (\ref{flow-jump-sets-out}) can be used to directly tune the minimum inter-transmission time (up to $\mathcal{T}$ in (\ref{T-phi})). It has to be noted that in this case, we can replace $\gamma^{2} W^{2}(e) \leq \delta(y)$ in (\ref{eq:tig-condn}) by $\gamma^{2} W^{2}(e) \leq (\alpha(|x|) + H^{2}(x) + \delta(x))$ when Assumption \ref{ass-clock-out} holds. The following result is a direct consequence of Theorem \ref{thm: clock-output}.
\begin{cor}
Suppose that Assumption \ref{ass-clock-out} holds and consider system (\ref{eq: hybrid-model}) with $y=x$ and the flow and jump sets defined as
\begin{equation}
\arraycolsep=0.8pt\def\arraystretch{1.3}
\begin{array}{lcll}
C &=& \Big\{q: &\gamma^{2} W^{2}(e) \leq \sigma(\alpha(|x|) + H^{2}(x) + \delta(x)) \text{ or } \tau \in [0,T]\Big\} \\[6pt]
D &=& \Big\{q: &\Big(\gamma^{2} W^{2}(e) = \sigma(\alpha(|x|) + H^{2}(x) + \delta(x)) \text{ and } \tau \geq T\Big) \text{ or } \\[6pt]
& & & \Big(\gamma^{2} W^{2}(e) \geq \sigma(\alpha(|x|) + H^{2}(x) + \delta(x)) \text{ and } \tau = T\Big) \Big\},
\end{array}
\end{equation}
where $q := (x,e,\tau)$, $\sigma \in (0,1)$ and $T$ is such that $T \in (0,\mathcal{T}(\gamma, L))$. Then, the conclusions of Theorem \ref{thm: clock-output} hold.
\hfill $\Box$
\end{cor}

\begin{exmple}
We illustrate the interest of our proposed triggering condition. Consider the LTI system, as in \cite{Tabuada2007event}, $\dot{x} = Ax + Bu$, where $x \in \R{2}{}$, $u\in \R{}{}, A = \left(\begin{smallmatrix}0 &\hspace{6pt} 1\\[6pt] -2 &\hspace{6pt} 3\end{smallmatrix}\right)$ and $B = \left(\begin{smallmatrix}0 \\[6pt] 1\end{smallmatrix}\right)$. Since the pair ($A,B$) is stabilizable, we take the control input $u=Kx$ with $K = [1 \,\,-4]$ as in \cite{Tabuada2007event}. By following similar lines as in Section \ref{sec: application-linear}, we derive the LMI (\ref{eq: prop-LMI}) with $\mathcal{A}_{1}=\mathcal{A}_{2}=A+BK$, $\mathcal{B}_{1}=\mathcal{B}_{2}=BK$ and $\varepsilon_{1}=0$. Hence, and by solving the resulted LMI, we obtain the numerical values $L = 4.1231, \varepsilon_{2} = 0.68, \gamma = 17.3495$ which lead to $\mathcal{T} = 0.079$. For comparison, we set $T = 0.075$ and we ran simulations for 200 randomly distributed initial conditions such that $|(x(0,0),e(0,0))|\leq 100$ and $\tau(0,0)=0$. Table \ref{tbl:tau-min-avg} provides the generated minimum and average inter-transmission times by both the proposed triggering strategy and the triggering condition in \cite{Tabuada2007event}, \textit{i.e.} with $T = 0$. We note that our proposed mechanism produces larger values of $\tau_{\min}$, $\tau_{\avg}$. To spotlight the effect of the time-triggered part in our proposed triggering mechanism, the enforced lower bound $T$ is plotted in Figures 2, 3 versus the generated inter-transmission times by both our proposed triggering mechanism and the triggering condition in \cite{Tabuada2007event} respectively, for one initial condition. \hfill $\Box$

\begin{table}[H]
\begin{center}
  \begin{tabular}{ l | c | c }
                & Our proposed triggering mechanism & Tabuada \cite{Tabuada2007event}  \\ \hline
  $\tau_{\min}$ & 0.075                                   & 0.0543 \\ \hline
  $\tau_{\avg}$ & 0.0772                                  & 0.0659
  \end{tabular}
  \caption{Minimum and average inter-execution times for 200 initial conditions such that $|(x(0,0),e(0,0))|\leq 100$ and $\tau(0,0)=0$ for a simulation time of 10 s.}
  \label{tbl:tau-min-avg}
\end{center}
\end{table}

\begin{figure}[H]
\centering
\psfrag{T}[][][1]{\scriptsize $\color{Red3}\textbf{T}\,\,\,\,$}
\includegraphics[width=\linewidth,height=4cm]{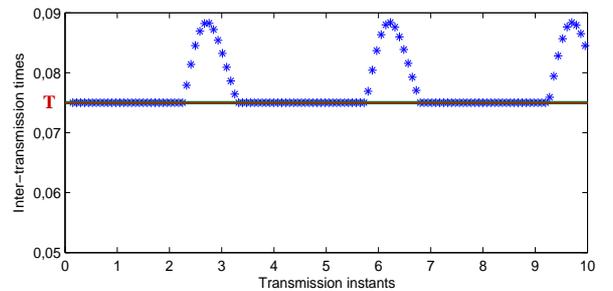}\caption{Inter-transmission times with $T = 0.075$.}
\label{fig:tau-state-clock}
\end{figure}

\begin{figure}[H]
\centering
\psfrag{T}[][][1]{\scriptsize $\color{Red3}\textbf{T}\,\,\,\,$}
\includegraphics[width=\linewidth,height=4cm]{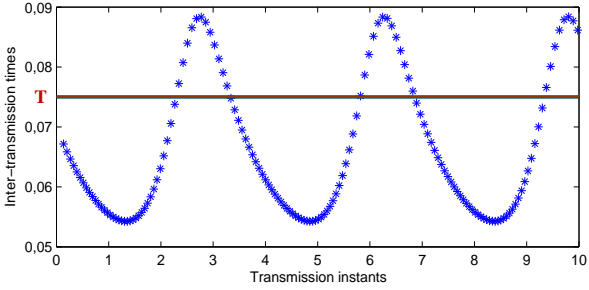} \caption{Inter-transmission times with \cite{Tabuada2007event}.}
\label{fig:tau-state-Tabuada}
\end{figure}
\end{exmple}

\section{Conclusion} \label{sec: conclusions}
We have developed output-based event-triggered controllers for the stabilization of nonlinear systems. The proposed technique ensures an asymptotic stability property and enforces a minimum amount of time between two consecutive transmission instants. The required conditions are shown to be satisfied by any stabilizable and detectable LTI systems. We show in \cite{Abdelrahim2014codesign} that these results can be used as a starting point to address the challenging co-design problem in which the output feedback law is not obtained by emulation but is jointly synthesized with the triggering condition.

\section*{Appendix}
\noindent\textbf{Proof of Theorem \ref{thm: clock-output}.}
First, we prove the result when Assumption \ref{ass-clock-out} holds globally. Let $\zeta: \R{}{\geq 0} \rightarrow \R{}{}$ be the solution to
\begin{equation} \label{zeta-dot}
  \dot{\zeta} = -2L\zeta - \lambda(\zeta^{2} + 1) \hspace{20pt} \zeta(0) = \theta^{-1},
\end{equation}
where $\theta \in (0,1)$, $\lambda := \sqrt{\gamma^{2} + \eta}$ for some $\eta > 0$ and $L, \gamma$ come from Assumption \ref{ass-clock-out}. We denote $\widetilde{\mathcal{T}}(\theta, \eta, \gamma, L)$ the time it takes for $\zeta$ to decrease from $\theta^{-1}$ to $\theta$. This time $\widetilde{\mathcal{T}}(\theta, \eta, \gamma, L)$ is a continuous function of $(\theta, \eta)$ which is decreasing in $\theta$ and $\eta$ (by invoking the comparison principle). In addition, it holds that $\widetilde{\mathcal{T}}(\theta, \eta, \gamma, L) \rightarrow \mathcal{T}(\gamma, L)$ as $(\theta, \eta)$ tends to $(0, 0)$ (where $\mathcal{T}(\gamma, L)$ is defined in Section \ref{sec: main-results}), like in \cite{Nesic2009explicit}. As a consequence, since $T< \mathcal{T}(\gamma, L)$, there exists $(\theta, \eta)$ such that $T<\widetilde{\mathcal{T}}(\theta, \eta, \gamma, L)$. We fix the couple $(\theta, \eta)$. Let $q := (x,e,\tau)$. We define for all $q \in C\cup D$, $R(q) := V(x) + \max\{0, \lambda\zeta(\tau)W^{2}(e)\}$. Let $q\in D$, we obtain, in view of (\ref{eq: hybrid-model}) and the fact that $W$ is positive definite,
\begin{align}
  R(G(q)) = V(x) + \max\{0, \lambda\zeta(0)W^{2}(0)\} = V(x) \leq R(q), \label{R-all-D}
\end{align}
where $G(q) := (x,0,0)$. Let $q\in C$ and suppose that $\zeta(\tau) < 0$. As a consequence it holds that $\tau > T$. Indeed, $\zeta(\tau)$ is strictly decreasing in $\tau$, in view of (\ref{zeta-dot}), and $\zeta(T)>\zeta(\widetilde{\mathcal{T}}(\theta, \eta, \gamma, L)) = \theta > 0$ as $T<\widetilde{\mathcal{T}}(\theta, \eta, \gamma, L)$. As a consequence $\zeta(\tau) < 0$ implies that $\tau>T$. Hence, $\gamma^{2} W^{2}(e) \leq  \delta(y)$ in view of (\ref{flow-jump-sets-out}) since $q \in C$. Consequently, in view of page 100 in \cite{Teel-Praly-mcss-00}, Lemma \ref{lma: clarke}, Assumption \ref{ass-clock-out} and the definition of the function $R$, $ R^{\circ}(q; F(q)) = V^{\circ}(x; f(x,e)) \leq -\alpha(|x|)$, where $F(q) := (f(x,e), g(x,e), 1)$. Hence, by following similar arguments as in the proof of Theorem 1 in \cite{Nesic2009explicit} since $\alpha$ is continuous and positive definite and $V$ is positive definite and radially unbounded, there exists a continuous positive definite function $\rho_{1}$ such that
\begin{align}
  R^{\circ}(q; F(q)) \leq -\rho_{1}(V(x)) = -\rho_{1}(R(q)). \label{Rdot1}
\end{align}
When $q \in C$ and $\zeta(\tau) > 0$, we have $R(q) = V(x) + \lambda\zeta(\tau)W^{2}(e)$. As above, in view of Lemma \ref{lma: clarke}, Assumption \ref{ass-clock-out} and (\ref{zeta-dot}) and by following the same lines as in the proof of Theorem 1 in \cite{Nesic2009explicit}, we obtain $R^{\circ}(q; F(q))  \leq  -\alpha(|x|)  -  H^{2}(x)  -  \delta(y)  +  \gamma^{2} W^{2}(e)  +  2\lambda \zeta(\tau)W(e)H(x)  -  \lambda^{2}\zeta^{2}(\tau)W^{2}(e)  -  \lambda^{2} W^{2}(e)$. Using the fact that $2\lambda \zeta(\tau)W(e)H(x) \leq \lambda^{2}\zeta^{2}(\tau)W^{2}(e) + H^{2}(x)$, $ R^{\circ}(q; F(q)) \leq -\alpha(|x|) -\delta(y) + \gamma^{2} W^{2}(e) - \lambda^{2} W^{2}(e) \leq -\alpha(|x|) + \gamma^{2} W^{2}(e) - \lambda^{2} W^{2}(e)$. Recall that $\lambda^{2} = \gamma^{2} + \eta$, it holds that $R^{\circ}(q; F(q)) \leq -\alpha(|x|) - \eta W^{2}(e)$. By using the same argument as in (\ref{Rdot1}), we derive that $R^{\circ}(q; F(q)) \leq -\rho_{1}(V(x)) - \eta W^{2}(e) = -\rho_{1}(V(x)) - \frac{\eta\theta}{\lambda} \lambda\theta^{-1}W^{2}(e) = -\rho_{1}(V(x)) - \rho_{2}(\lambda\theta^{-1}W^{2}(e))$, where $\rho_{2}: s \mapsto \frac{\eta\theta}{\lambda}s \in \Kinf$. Since $\zeta(\tau) \leq \theta^{-1}$ for all $\tau \geq 0$ in view of (\ref{zeta-dot}), it holds that $R^{\circ}(q; F(q)) \leq -\rho_{1}(V(x)) - \rho_{2}(\lambda\zeta(\tau)W^{2}(e))$. We deduce that there exists a continuous positive definite function $\rho_{3}$ such that $R^{\circ}(q; F(q)) \leq -\rho_{3}(V(x) + \lambda\zeta(\tau)W^{2}(e)) = - \rho_{3}(R(q))$. In view of the last inequality, (\ref{Rdot1}) and Lemma \ref{lma: clarke}, when $\zeta(\tau)=0, R^{\circ}(q; F(q)) \leq \max\{-\rho_{1}(R(q)), -\rho_{3}(R(q))\}$. Consequently, it holds that, for all $q\in C$
\begin{equation} \label{R-all-clark}
  R^{\circ}(q; F(q)) \leq -\rho(R(q))
\end{equation}
where $\rho := \min\{\rho_{1}, \rho_{3}\}$ is continuous and positive definite. Let $\phi$ be a solution to (\ref{eq: hybrid-model}), (\ref{flow-jump-sets-out}). In view of (\ref{R-all-clark}) and by definition of the Clarke's derivative (see for instance page 99 in \cite{Teel-Praly-mcss-00}), it holds that, for all $j$ and for almost all $t\in I^{j}$ (where $I^{j} = \{ t: (t,j)\in\dom\phi \}$)
\begin{equation} \label{R-all-C}
  \dot{R}(\phi(t,j)) \leq R^{\circ}(\phi(t,j); F(\phi(t,j))) \leq -\rho(R(\phi(t,j))).
\end{equation}
Thus, in view of (\ref{R-all-D}), (\ref{R-all-C}) and since inter-jump times are lower bounded by $T$ in view of (\ref{flow-jump-sets-out}), we conclude that, by following the same lines as in the end of the proof of Theorem 1 in \cite{Nesic2009explicit}, there exists $\tilde{\beta} \in \KL$ such that for any solution $\phi$ to (\ref{eq: hybrid-model}), (\ref{flow-jump-sets-out}) and any $(t,j) \in \dom \phi$, $R(\phi(t,j)) \leq \tilde{\beta}(R(\phi(0,0)), 0.5t+0.5Tj)$. In view of Assumption \ref{ass-clock-out} and since $W$ is continuous (since it is locally Lipschitz) and positive definite, there exists $\overline{\alpha}_{W}\in \Kinf$ such that $W(e) \leq \overline{\alpha}_{W}(|e|)$ for all $e \in \R{n_{e}}{}$ according to Lemma 4.3 in \cite{Khalil}. As a result, in view of Assumption \ref{ass-clock-out}, (\ref{zeta-dot}) and the definition of the function $R$, it holds that, for all $q \in C\cup D$, $\underline{\alpha}(|x|) \leq R(q) \leq \overline{\alpha}(|x|) + \frac{\lambda}{\theta}\overline{\alpha}_{W}(|e|) \leq \overline{\alpha}_{R}(|(x,e)|)$, where $\overline{\alpha}_{R}: s \mapsto \overline{\alpha}(s) + \frac{\lambda}{\theta}\overline{\alpha}_{W}(s) \in \Kinf$. Hence, we deduce that (\ref{eq-thm-clock-out}) holds for any solution $\phi$ to (\ref{eq: hybrid-model}), (\ref{flow-jump-sets-out}) ad for all $(t,j) \in \dom \phi$, where $\beta: (s_{1}, s_{2}) \mapsto \underline{\alpha}^{-1}(\tilde{\beta}(\overline{\alpha}_{R}(s_{1}),s_{2})) \in \KL$.

We now investigate the completeness of the maximal solutions to system (\ref{eq: hybrid-model}), (\ref{flow-jump-sets-out}). Let $\phi$ be a maximal solution to (\ref{eq: hybrid-model}), (\ref{flow-jump-sets-out}). We first show that $\phi$ is nontrivial, \textit{i.e.} its domain contains at least two points (see Definition 2.5 in \cite{Teel}). According to Proposition 6.10 in \cite{Teel}, it suffices for that purpose to prove that $\{F(q)\}\cap T_{C}(q) \neq \emptyset$ for any $q := (x,e,\tau)\in C\backslash D$, where $F(q) := (f(x,e), g(x,e), 1)$ and $T_{C}(q)$ is the tangent cone\footnote{The tangent cone to a set $S\subset \R{n}{}$ at a point $x\in\R{n}{}$, denoted $T_{S}(x)$, is the set of all vectors $\omega\in\R{n}{}$ for which there exist $x_{i}\in S, \tau_{i}>0$ with $x_{i}\rightarrow x, \tau \rightarrow 0$ as $i \rightarrow \infty$ such that $\omega = \lim_{i\rightarrow \infty} (x_{i} - x)/\tau_{i}$ (see Definition 5.12 in \cite{Teel}).} to $C$ at $q$. Let $q\in C\backslash D$. If $q$ is in the interior of $C$, $T_{C}(q) = \R{n_{x} + n_{e} + 1}{}$ and the required condition holds. If $q$ is not in the interior of $C$, necessarily $\tau = 0$ as $q\in C\backslash D$, in this case $T_{C}(q) = \R{n_{x}+ n_{e}}{}\times \R{}{\geq0}$ and we see that $F(q) \in T_{C}(q)$, in view of (\ref{eq: hybrid-model}). Hence, $\phi$ is nontrivial according to Proposition 6.10 in \cite{Teel}. In view of (\ref{eq: hybrid-model}), (\ref{flow-jump-sets-out}) and (\ref{eq-thm-clock-out}), $\phi_{x}$ and $\phi_{\tau}$ cannot explode in finite time. Recall that the network-induced error is  $\phi_{e} = (\phi_{e_{y}}, \phi_{e_{u}})$ with $\phi_{e_{y}} = \phi_{y}(t_{j},j) - \phi_{y}(t,j)$, $\phi_{e_{u}} = \phi_{u}(t_{j},j) - \phi_{u}(t,j)$ for $j > 0$ and $(t,j)\in \dom \phi$ where we write $\dom \phi = \cup_{j\in\{0,...,J\}}([t_{j}, t_{j+1}], j)$ with $J\in\Zp\cup\{\infty\}$. Hence, in view of (\ref{eq: xdot_p-y}), (\ref{eq: xdot_c-u}), (\ref{eq-thm-clock-out}) and since $g_{p}$, $g_{c}$ are continuous, it holds that, for all $j > 0$ and $(t,j) \in \dom \phi$,
\begin{equation}\label{ey-bound}
\begin{array}{lcl}
  |\phi_{e_{y}}(t,j)| &=& |g_{p}(\phi_{x_{p}}(t_{j},j)) - g_{p}(\phi_{x_{p}}(t,j))| \\[6pt]
                      &\leq& |g_{p}(\phi_{x_{p}}(t_{j},j))| + |g_{p}(\phi_{x_{p}}(t,j))| \\[6pt]
                      &\leq& \underset{|z|\leq\beta(|(\phi_{x}(0,0), \phi_{e}(0,0))|, 0)}{2\hspace{12pt}\max|g_{p}(z)|.}
\end{array}
\end{equation}
Similarly, we obtain, for all $j > 0$ and $(t,j) \in \dom \phi$
\begin{equation}\label{eu-bound}
\begin{array}{lcl}
  |\phi_{e_{u}}(t,j)| &\leq& |g_{c}(\phi_{x_{c}}(t_{j},j), g_{p}(\phi_{x_{p}}(t_{j},j))| \\[6pt]
                      &\quad& + |g_{c}(\phi_{x_{c}}(t,j), g_{p}(\phi_{x_{p}}(t_{j},j))| \\[6pt]
                      &\leq& \underset{\begin{subarray}{l}|z_{1}|\,\leq \,\beta(|(\phi_{x}(0,0), \phi_{e}(0,0))|, 0)\\[3pt]
                                                         |z_{2}|\,\leq \,\max|g_{p}(z_{1})|\end{subarray}}{2\hspace{12pt}\max|g_{c}(z_{1},z_{2})|.}
\end{array}
\end{equation}
When $j = 0$, we have that $|\phi_{e_{y}}(t,0)| \leq |\phi_{e_{y}}(0,0)| + |g_{p}(\phi_{x_{p}}(0,0)) - g_{p}(\phi_{x_{p}}(t,0))|$ and $|\phi_{e_{u}}(t,0)| \leq |\phi_{e_{u}}(0,0)| + |g_{c}(\phi_{x_{c}}(0,0), \phi_{y}(0,0)) - g_{c}(\phi_{x_{c}}(t,0), \phi_{y}(0,0))|$ and we can derive similar bounds on the interval $[0,t_{1}]$. As a result, and since $\phi_{e}$ is reset to 0 at each jump, $\phi_{e}$ cannot blow up in finite time. As a consequence, $\phi$ cannot explode in finite time. Let $G(x,e,\tau) := (x,0,0)$ denotes the jump map in (\ref{flow-jump-sets-out}). The solutions to (\ref{eq: hybrid-model}), (\ref{flow-jump-sets-out}) cannot leave the set $C\cup D$ after a jump since $G(D) \subset C$ in view of (\ref{eq: hybrid-model}), (\ref{flow-jump-sets-out}). Thus, we conclude that maximal solutions to (\ref{eq: hybrid-model}), (\ref{flow-jump-sets-out}) are complete according to Proposition 6.10 in \cite{Teel}. Finally, we note that if Assumption \ref{ass-clock-out} holds locally, then there exists $\Delta>0$ such that (\ref{R-all-D}) and (\ref{R-all-C}) hold on the invariant set $|(x,e)|\leq \Delta$ and consequently (\ref{eq-thm-clock-out}) holds locally.  \hfill $\Box$


\vspace{0.5cm}
\noindent \textbf{Proof of Proposition \ref{prop: linear-systems}.}
Let $W(e) = |e|$. Then, in view of (\ref{eq: hybrid-model-linear}), we have that, for all $x \in \R{n_{x}}{}$ and almost all $e \in \R{n_{e}}{}$
\begin{equation}
  \langle\nabla W(e), \mathcal{A}_{2}x + \mathcal{B}_{2}e\rangle \leq |\mathcal{A}_{2}x| + |\mathcal{B}_{2}||e|.
\end{equation}
Hence, condition (\ref{W-dot-out}) holds with $L = |\mathcal{B}_{2}|$ and $H(x) = |\mathcal{A}_{2}x|$. Let $V(x) = x^{T}Px$, where $P$ is real positive definite and symmetric. Therefore, condition (\ref{Vx-bounds-out}) is satisfied with $\underline{\alpha}(|x|) = \lambda_{\min}(P)|x|^{2}$ and $\overline{\alpha}(|x|) = \lambda_{\max}(P)|x|^{2}$. Consequently, for all $e \in \R{n_{e}}{}$ and almost all $x \in \R{n_{x}}{}$
\begin{equation} \label{vx-dot}
\begin{array}{lcl}
  \langle\nabla V(x), \mathcal{A}_{1}x + \mathcal{B}_{1}e\rangle &=& x^{T}(\mathcal{A}_{1}^{T}P + P\mathcal{A}_{1})x + x^{T}P\mathcal{B}_{1}e \\[2pt]
                                                        & &+ e^{T}\mathcal{B}_{1}^{T}Px.
\end{array}
\end{equation}
By post- and pre-multiplying LMI (\ref{eq: prop-LMI}) respectively by the state vector $(x,e)$ and its transpose, we obtain
\begin{equation}\small
\arraycolsep=1pt\def\arraystretch{1.3}
\begin{array}{rrrrr}
  \left(\begin{array}{c} x \\ e \end{array}\right)^{T}\!\left(\begin{array}{cc} \mathcal{A}_{1}^{T}P\!+\! P\mathcal{A}_{1}\!  + \! \mathcal{A}_{2}^{T}\mathcal{A}_{2}\! + \!\varepsilon_{1}\overline{C}_{p}^{T}\overline{C}_{p} \! + \!\varepsilon_{2}\mathbb{I}_{n_{x}} \!&\!P\mathcal{B}_{1} \\ \mathcal{B}_{1}^{T}P\!&\!-\mu\mathbb{I}_{n_{e}} \end{array}\right)\!\left(\begin{array}{c} x \\ e \end{array}\right) \leq 0.
\end{array}
\end{equation}
Expanding the last inequality yields
\begin{equation}\label{eq: proof-prop-lmi}
\begin{array}{rrrrr}
  x^{T}(\mathcal{A}_{1}^{T}P + P\mathcal{A}_{1})x + x^{T}P\mathcal{B}_{1}e + e^{T}\mathcal{B}_{1}^{T}Px \leq -\varepsilon_{2} x^{T}x \\[2pt]
  - x^{T}\mathcal{A}_{2}^{T}\mathcal{A}_{2}x - \varepsilon_{1}x^{T}\overline{C}_{p}^{T}\overline{C}_{p}x + \mu e^{T}e
\end{array}
\end{equation}
which implies that
\begin{equation}\label{eq: proof-prop-lmi}
\begin{array}{rrrrr}
  x^{T}(\mathcal{A}_{1}^{T}P + P\mathcal{A}_{1})x + x^{T}P\mathcal{B}_{1}e + e^{T}\mathcal{B}_{1}^{T}Px \leq -\varepsilon_{2}|x|^{2} \\[2pt]
  - |\mathcal{A}_{2}x|^{2} - \varepsilon_{1}|\overline{C}_{p}x|^{2} + \mu|e|^{2}.
\end{array}
\end{equation}
As a result, in view of (\ref{vx-dot}), (\ref{eq: proof-prop-lmi}), condition (\ref{V-dot-out}) is verified with $\alpha(|x|)=\varepsilon_{2}|x|^{2}$, $\delta(y)=\varepsilon_{1}|y|^{2}$ and $\gamma = \sqrt{\mu}$. Thus, Assumption \ref{ass-clock-out} holds. \hfill $\Box$


\bibliographystyle{IEEEtran}
\bibliography{D:/Mahmoud/References}
\end{document}